\documentclass[pra,twocolumn,showpacs,showkeys,secnumarabic,amssymb,amsmath,superscriptaddress,floatfix]{revtex4-1}
\usepackage{amssymb,amsmath}
\usepackage{color}
\usepackage{tikz}
\usepackage{mathtools}
\usetikzlibrary{arrows.meta}
\usepackage{graphics}
\usepackage{graphicx}
\usepackage{enumitem}
\usepackage{bm}
\begin{document}

\title{Basis-independent quantum coherence and its distribution}
\author{Chandrashekar Radhakrishnan}
\affiliation{New York University Shanghai, 1555 Century Ave, Pudong, Shanghai, 200122, China.}

\author{Zhe Ding}
\affiliation{CAS Key Laboratory of Microscale Magnetic Resonance and Department of Modern Physics, University of Science and Technology of China, Hefei 230026, China}

\author{Fazhan Shi}
\affiliation{CAS Key Laboratory of Microscale Magnetic Resonance and Department of Modern Physics, University of Science and Technology of China, Hefei 230026, China}

\author{Jiangfeng Du}
\affiliation{CAS Key Laboratory of Microscale Magnetic Resonance and Department of Modern Physics, University of Science and Technology of China, Hefei 230026, China}

\author{Tim Byrnes}
\email{tim.byrnes@nyu.edu}
\affiliation{New York University Shanghai, 1555 Century Ave, Pudong, Shanghai, 200122, China.}
\affiliation{Department of Physics, New York University, New York, NY, 10003, USA.}

\begin{abstract}
We analyze a basis-independent definition of quantum coherence.  The maximally mixed state is used as the reference state, which allows for a way of defining coherence that is invariant under arbitrary unitary transformations.  The basis-independent approach 
is applied to finding the distribution of the coherence within a multipartite system, where the contributions due to correlations between the subsystems and within each subsystem are isolated.  The use of the square root of the Jensen-Shannon divergence allows for inequality relations to be derived between these quantities, giving a geometrical picture within the Hilbert space of the system. We describe the relationship 
between the basis-independent and the basis-dependent approaches, and argue that 
many advantages exist for the former method.  The formalism is illustrated with several numerical examples which show that the states can be characterized in a simple and effective manner.   
\end{abstract}


\date{\today}

\maketitle

\section{Introduction}
\label{intro}
Coherence has always been one of the central concepts in quantum mechanics distinguishing it from classical physics.  In the field of quantum 
optics it was understood in the context of phase space distributions \cite{glauber1963coherent} 
and higher order correlation functions \cite{sudarshan1963equivalence,scully1999quantum}.
Although these approaches gave a basic understanding of the nature of coherence, it was not quantified in a rigorous sense.  In the seminal work of Baumgratz, Cramer, and Plenio \cite{baumgratz2014quantifying}, this was achieved by defining central concepts such as incoherent states, maximally coherent states and incoherent operations.  In a short time rapid developments have been 
made in the basic theory of quantum coherence \cite{du2015conditions,ma2016converting,
cheng2015complementarity,lostaglio2015quantum,yao2015quantum,zhang2016quantifying,zhang2017quantum,
xu2016quantifying} 
and its applications \cite{opanchuk2016quantifying,zheng2016detecting,man2015cavity,mani2015cohering,yadin2016general}.  
On the application side the major focus has been on using quantum coherence for investigating quantum 
state merging \cite{streltsov2016entanglement}, in assisted subspace 
discrimination \cite{zhang2017intrinsic} and identifying quantum phase transitions 
\cite{karpat2014quantum,malvezzi2016quantum,cheng2018multipartite,sha2018thermal,
radhakrishnan2017quantum,radhakrishnan2017quantum2} in many body systems. Meanwhile, investigations into the basic theory of quantum coherence
have mainly focused on introducing new measures of quantum coherence 
\cite{baumgratz2014quantifying,shao2015fidelity,rana2016trace,radhakrishnan2016distribution,
napoli2016robustness,yao2016frobenius,rastegin2016quantum,girolami2014observable,yuan2015intrinsic,
yu2016alternative} and understanding the distribution of quantum coherence in multipartite coherence 
\cite{radhakrishnan2016distribution,radhakrishnan2016coherence}.

Quantum resource theories investigate the limits of performance for quantum information tasks, and 
also give bounds to  
potential applications based on the resource.  The resource theory
of entanglement is well-developed and have found applications in a wide variety 
of quantum technological applications \cite{horodecki2013quantumness,giovannetti2011advances,giovannetti2006quantum,escher2011general}.  
The development of the resource theory of quantum coherence is currently a very active field of 
research \cite{winter2016operational,yadin2016quantum,chitambar2016critical,streltsov2017colloquium,
yang2018operational,chitambar2016relating,streltsov2017structure}.  To develop this resource theory and relate them to the existing resource theories a critical understanding of the fundamental properties of quantum coherence is essential.  Quantum coherence has several distinctive features that contrast it to other properties of quantum states such as entanglement.  For instance, 
coherence can exist localized within an individual qubit and also as correlations 
between the qubits \cite{radhakrishnan2016distribution}.  Furthermore, according to the original formulation of Ref. \cite{baumgratz2014quantifying}, it is a basis-dependent property.  Of these properties, the first is a fundamental nature of quantum coherence, while the second arises due to the definition of what an ``incoherent state'' should be.

Since the introduction of the quantum theory of coherence, the basis-dependent aspect of coherence has been a point of controversy and confusion.  Other quantities characterizing quantum states (e.g. entanglement, discord, etc.) generally are basis-independent, and as such can be viewed as an objective property that the state possesses.  A basis-dependent quantity brings a level of subjectivity to coherence, as it is only quantified relative to a particular convention chosen by the observer.  To overcome this point, several methods such as
optimizing over all possible bases \cite{yu2017quantum,luo2017quantum} and 
using the maximally mixed state as the incoherent state \cite{yu2016total,yao2016frobenius,hu2017maximum,zhang2017classical} have been proposed.
Currently there is no consensus as to which is the more appropriate method of measuring coherence, and when it is suitable to use basis-dependent and basis-independent methods. In order that the theory quantum coherence can be utilized in a useful way, it is an important question to clarify the difference between the existing approaches.  

In this paper, we analyze a basis-independent approach of measuring quantum coherence, extending the basis-dependent approach of Ref.  \cite{baumgratz2014quantifying}.  The approach has a somewhat different interpretation to the quantity being measured, and more closely related to the purity of the system, rather than the superposition in a particular basis.  We argue that this still quantifies coherence in that any state that is not the maximally mixed state has coherence in some basis.  We also investigate alternative methods of breaking down 
the distribution of coherence in a multipartite system.  In our previous work \cite{radhakrishnan2016distribution}, we decomposed the total coherence into two parts, the intrinsic and local coherence, based on separable states.  Here we analyze another decomposition based on product states which finds the coherence due to correlations between subsystems, and those which are completely localized.  These are called the collective and localized coherences respectively, and are shown to have several favorable properties when characterizing the coherence distribution.  In this work, we discuses primarily a measure based on the square root of the Jensen-Shannon divergence, which has several mathematically favorable properties.  This measure is the most appropriate in terms of decomposition of coherence, primarily because it satisfies a triangle inequality and is a metric,  However, we expect that similar results can be obtained with other measures, as has been 
performed already in several works \cite{kumar2017quantum,tan2016unified,radhakrishnan2016distribution}.
We have therefore kept the definition of the various decompositions in general form such that they  be applied to different measures, and specifying the Jensen-Shannon divergence as our particular case.

The outline of this article is as follows.   A brief introduction of the
various statistical divergence measures is given in Sec. \ref{jsd}, in particular we 
explain the Jensen-Shannon divergence.  In Sec. \ref{basisindependent} we introduce a 
basis-independent measure of quantum coherence and discuss its properties.  The application of the basis-independent approach to 
finding the distribution of coherence in multipartite systems is discussed in Sec. \ref{sec:dist}, 
where the various contributions are identified and the geometrical connections are discussed.  In 
Sec. \ref{qcmpml} we illustrate our methods taking the example of the two-site transverse Ising model, the GHZ and W class of states.  Finally, in Sec. \ref{sumconc} we summarize and present our conclusions.

\section{The Jensen-Shannon divergence}
\label{jsd}

We give a brief review of the properties of the Jensen-Shannon divergence. While this
is a well-known quantity in classical probability theory, it has a relatively short history in the context of quantum information theory \cite{majtey2005jensen,majtey2008jensen,kenfack2018dynamics}.  In this section will discuss its properties in classical and quantum context.  This will serve 
to introduce our notation and summarize the relevant properties for applications to quantum coherence.

\subsection{Classical properties} 

Comparing two probability distributions $P = \{p_{1},...,p_{m}\}$ and 
$Q = \{q_{1},...,q_{m}\}$ is a fundamental task that is relevant to many applications.  One possible 
method is to use the normed distance between these distributions.  For example the $\ell_{n}$-norm
between these two distributions is defined as 
\begin{equation}
\ell_n(P,Q) = \left( \sum_{i=1}^{m} |p_{i} - q_{i}|^{n} \right)^{1/n},
\end{equation}
where for $n=1$ we have the $\ell_{1}$-norm and $n=2$ is the well known Euclidean norm.  
Another option is to use a contrast function which compares the information in one distribution 
to another.  This can be achieved through the use of relative entropy 
\begin{equation}
S(P \| Q) = \sum_{i=1}^{m} p_{i} \log \frac{p_{i}}{q_{i}},
\end{equation}
which estimates the entropic difference between the two distributions. 

Each of these methods have advantages and disadvantages that make them more or less suitable depending upon the application.  
Both quantities are relatively simple to evaluate, but in quantum information contexts it is often desirable to have an entropic 
quantity such that there is a direct comparison of the information contained in two probability 
distributions. However, the relative entropy has 
two major disadvantages:  it is asymmetric in the probability distributions and hence cannot be used 
as a distance in the statistical manifold. For a measure ${\cal D}$ to be called a distance over a space $X$, 
it should satisfy the following axioms: 
(i) ${\cal D}(x,y) \geq 0 \quad \forall \; \; x \neq y$ with ${\cal D}(x,x) = 0$. 
(ii) ${\cal D}(x,y) = {\cal D}(y,x)$. 
In addition if the function verifies the triangle inequality ${\cal D}(x,y) + {\cal D}(y,z) \geq {\cal D}(x,z)$ then 
${\cal D}$ is also a metric for the space $X$. If a measure obeys the positivity axiom as in 
(i) then it is generally referred to as a divergence.   
Since the relative entropy is not a distance the measured 
value quantifies the entropic deviation from $Q$ to $P$ and not the other way round.  Further it 
requires that the set of non-zero probabilities in the set $Q$ should be larger than or equal to 
the corresponding ones in the set $P$. This above problem is referred to as the support problem of the 
relative entropy \cite{lin1991divergence}, where the term support refers to the set of non-trivial probabilities.  

To overcome the disadvantages of the relative entropy, two different functions namely 
the $\mathsf{J}$ and $\mathsf{S}$ divergences were defined in Ref. \cite{lin1991divergence}
\begin{eqnarray}
\mathsf{J}(P,Q) &=& \frac{1}{2} \left(S(P \| Q) + S(Q \| P)\right), \label{jdivergence} \\
\mathsf{S}(P,Q) &=&  P \log \left( \frac{2P}{P + Q} \right).
\label{sdivergence}
\end{eqnarray}
Although the $\mathsf{J}$ divergence in equation (\ref{jdivergence}) is symmetric
and satisfies the axioms of distance, it places a more severe 
restriction that the support of $P$ and $Q$ must be equal.  Meanwhile, the $\mathsf{S}$-divergence in 
(\ref{sdivergence}) is well-defined irrespective of the support of $P$ and $Q$, but it is not symmetric
and hence not a distance. 
Incorporating the core idea behind these two functions, one can introduce the Jensen-Shannon 
divergence (JSD) \cite{lin1991divergence} as: 
\begin{equation}
J(P,Q) = \frac{1}{2} \left(\mathsf{S}(P,Q) + \mathsf{S}(Q,P)\right).
\end{equation}
In terms of entropy, the above equation can be written as
\begin{equation}
J(P,Q) = S\left(\frac{P+Q}{2}\right) - \frac{S(P)+ S(Q)}{2}
\end{equation}
Here $S(P) = - \sum_{i} p_{i} \log p_{i}$ is the Shannon entropy of the system.  
The JSD satisfies the axioms of distance but it does not obey the triangle inequality.  
This means that it is a distance, but not a metric.  The square root of JSD obeys the distance axioms
as well as the triangle inequality and hence qualifies as a metric
\cite{endres2003new}.

\subsection{Quantum properties} 

Similar comparisons can be made to quantum mechanical systems. The primary difference in the quantum case is that instead of probability distributions, we wish to compare two quantum states, written as density matrices $\rho,\sigma \; \in 
\mathcal{B}\left(\mathcal{H}_{1}^{+}\right)$.  The quantum version of the 
relative entropy is defined as \cite{umegaki1962conditional} 
\begin{equation}
S(\rho \| \sigma) = {\rm Tr} \left(\rho \log \rho - \rho \log \sigma \right).
\end{equation}
The quantum relative entropy also has the disadvantages of being asymmetric and is well defined only when
the support of $\sigma$ is equal to or larger than that of $\rho$.  Here the support is defined as the 
subspace spanned by the eigenvectors corresponding to the non-zero eigenvalues of an operator. 
Following the same procedure as in the classical scenario, we can overcome these problems by defining 
a quantum version of the Jensen-Shannon divergence (QJSD) \cite{briet2009properties}:
\begin{eqnarray}
\mathcal{J}(\rho,\sigma) &=& \frac{1}{2} \left[ S\left( \rho \Big| \Big| \frac{\rho+\sigma}{2}  \right) + 
              S\left( \sigma \Big| \Big|  \frac{\rho+\sigma}{2} \right) \right],\;
              \label{Jensen1} \\
			&=& S \left(\frac{\rho + \sigma}{2} \right) - \frac{S(\rho)+ S(\sigma)}{2} ,
			  \label{Jensen2}
\end{eqnarray}
where $S(\rho) = - {\rm Tr} \rho \log \rho$ is the von-Neumann entropy.  The QJSD is symmetric and 
well-defined irrespective of the nature of the support of $\rho$ and $\sigma$ \cite{majtey2005jensen}
and hence can be used as a measure of distinguishability between quantum states. 

Although the QJSD
is a distance, it is itself does not satisfy the triangle inequality and hence is not a metric.  On the other 
hand the square root of QJSD 
\begin{align}
{\cal D} (\rho,\sigma) = \sqrt{\mathcal{J}(\rho,\sigma)}
\label{qjsd}
\end{align}
is a distance measure and also obeys the triangle inequality for pure
states.  
For mixed states it has been conjectured that the square root of QJSD obeys the triangle
inequality and has been verified numerically \cite{lamberti2008metric}.   Hence 
$\left(\mathcal{B}(\mathcal{H}_{1}^{+}), \sqrt{\mathcal{J}}\right)$ is a metric space 
which can be isometrically embedded in a real Hilbert space. 

 In this paper, we will henceforth use the QJSD 
as our distance measure, due to the favorable properties it possesses as described above. Although we mainly have the QJSD in mind when defining various coherence quantities, we have included the more general expressions in terms of $ \cal D $ where appropriate such that they can also be applied to other distance measures.   It is important to note that different measures have different ordering of quantum states in the Hilbert space 
\cite{virmani2000ordering,okrasa2012two,liu2016ordering}.   While the same relationships do not always hold for all measures, in particular the triangle inequality, analogous properties are likely to be present for most measures.

\section{basis-independent measure of quantum coherence}
\label{basisindependent}

We now turn to applying the distance measures of the previous section to quantifying quantum coherence.  In the original 
formulation of Ref. \cite{baumgratz2014quantifying}, coherence is inherently a basis dependent quantity. While several 
measures have been used to quantify coherence ranging from relative entropy to $\ell_{1}$-norm, we have previously proposed 
the use of the square root of the QJSD \cite{radhakrishnan2016distribution,radhakrishnan2016coherence,radhakrishnan2017quantum,radhakrishnan2017quantum2}.  For a given quantum state $ \rho $, the coherence is defined as 
\begin{eqnarray}
C^{(b)} (\rho)  &\equiv& \min_{\sigma \in \mathcal{I}^{(b)}} {\cal D} (\rho,\sigma) , \\
         & = & \min_{\sigma \in \mathcal{I}^{(b)}} 
             \sqrt{S \left(\frac{\rho + \sigma}{2} \right) - \frac{S(\rho) + S(\sigma)}{2}},
\label{coherencedefinition}             
\end{eqnarray}
where $ \mathcal{I}^{(b)} $ is the set of all incoherent states.  In the second line we have written the explicit expression for the case (\ref{qjsd}) of using QSJD as the measure.   The minimization is performed such that $\sigma$ is the closest incoherent state (diagonal state) in a particular basis $ b $.  Since the set of incoherent states is dependent upon a particular choice of basis $ b $, the coherence is a basis dependent quantity. Our notation throughout this paper will be that if a coherence $ C $ has the label $^{(b)} $ it is a basis-dependent quantity, and otherwise it is basis-independent.  We note that the basis dependence does not arise from the use of QJSD or any choice of distance measure.  It arises due to the minimization over a particular set of incoherent states.

To overcome the basis dependence of quantum coherence, one alternative that has been suggested has been to use the maximally mixed state \cite{yu2016total,yao2016frobenius,hu2017maximum,zhang2017classical} as the reference state
\begin{equation}
\rho_{I} \equiv \frac{I}{d} = \frac{1}{d} \sum_{i=1}^{d} |i \rangle\langle i|,
\label{incoherentstate}
\end{equation}
where $I$ is the identity matrix in a $ d $ dimensional Hilbert space. It is well known that the 
state $\rho_{I}$
is a basis invariant state, i.e. it maintains the same form 
in any given basis.  Furthermore, it is unique in the sense that it is the only state where a unitary transformation maintains a diagonal structure with zero off-diagonal elements. In this case, $ \mathcal{I}^{(b)} $ only contains the state (\ref{incoherentstate}) and a basis-independent measure of quantum coherence is therefore given by 
\begin{align}
C(\rho) & \equiv  {\cal D} (\rho,\rho_I), \\
& = \sqrt{S \left( \frac{\rho+I/d}{2}\right) - \frac{S(\rho)+  \log_{2} d}{2} }.
\label{coherencemeasure}
\end{align}
This is equivalent to $ \mathcal{I}^{(b)} $ only containing one element of $ \rho_{I} $, such that the minimization does not need to be performed. 

The coherence measure (\ref{coherencemeasure}) has the important property that it is invariant under a general unitary transformation.   Consider performing an arbitrary unitary transformation of the state $ \rho \rightarrow U^\dagger \rho U $.  The coherence after the transformation is
\begin{align}
& C(U^\dagger \rho U) \nonumber \\
& =  \sqrt{S \left( \frac{ U^\dagger \rho U +I/d}{2} \right) - \frac{S(U^\dagger \rho U) + \log_{2} d}{2} }, \nonumber \\
& =  \sqrt{S \left( U^\dagger \frac{(\rho +I/d)}{2} U \right) - \frac{ S(\rho) + \log_{2} d}{2} }, \nonumber \\
& = C( \rho ) ,
\label{generalcoherence}               
\end{align}
which takes the same value before the basis transformation, since the entropy is only dependent on the 
eigenvalues of the density matrix.  This property ensures that the coherence is an absolute and not a 
relative quantity.  The basis-dependent definition is a relative property because it is dependent upon a 
particular choice of basis.  

For simple classes of states it is possible to have a closed expression for the coherence. For the Werner states
\begin{equation}
\rho (\mu) = \frac{(1 - \mu)}{d} I + \mu |\psi \rangle \langle \psi|,
\label{wernertype}
\end{equation}
where $0 \le \mu \le 1 $ is the mixing parameter, the coherence can be evaluated to be
\begin{align}
C(\rho (\mu) ) = & \frac{1}{2d} \Big[ (1+ \mu(d-1)) \log_2 (1+ \mu (d-1)) \nonumber \\
& - (2+ \mu(d-1)) \log_2 (1+ \frac{\mu}{2} (d-1)) \nonumber \\
& + (1-\mu)(d-1) \log_2 (1-\mu)  \nonumber \\
& - (2-\mu)(d-1) \log_2 (1- \frac{\mu}{2} ) \Big]^{1/2}.
\label{generalcoherenceformula}
\end{align}
In the limit of pure states  $ \mu = 1 $, this takes the value
\begin{align}
C(| \psi \rangle \langle \psi | )= \sqrt{1 + \frac{1}{2} \left( \log_2 d - (1+ \frac{1}{d}) \log_2 (d+1) \right) }.
\label{totalcoherenceformula}
\end{align}

What is the relationship between the basis-dependent definition of coherence as in (\ref{coherencedefinition}) and the basis-independent definition (\ref{coherencemeasure})?  In the original work of Ref. \cite{baumgratz2014quantifying}, the concept of coherence was closely related to the {\it superposition} principle.  This is clearly a basis-dependent concept as an arbitrary density matrix can always be diagonalized, giving zero off-diagonal elements and hence zero coherence. Thus in the approach of Ref. \cite{baumgratz2014quantifying}, the coherence is a measure of quantum superposition, which gives rise to a basis-dependent quantity. Meanwhile, in a definition of coherence such as (\ref{coherencemeasure}), the reference is a maximally mixed state $ \rho_{I}  $. In this case, 
the measure is more related to the {\it purity} of the system. Purity is a concept which is also closely related to coherence in the following sense.  Given any state that is not the completely mixed state $ \rho_{I}  $, one can always find a basis where there is a superposition of states, which contains off-diagonal elements to the density matrix.  On the other hand, $ \rho_{I}  $ cannot be put in a superposition state regardless of the basis chosen.

\begin{figure}[t]
\includegraphics[width=\columnwidth]{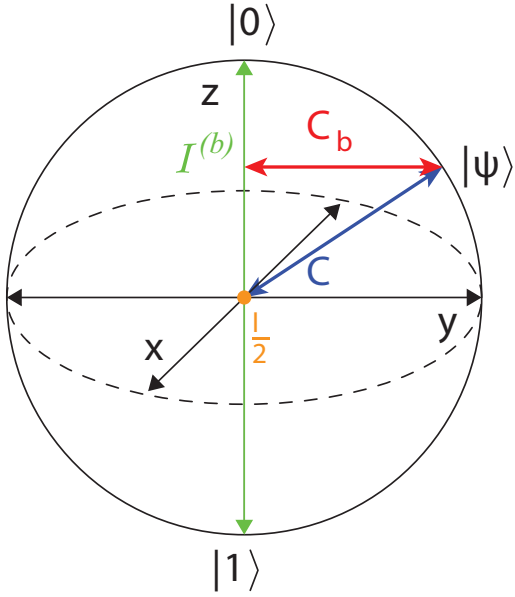}
\caption{Comparison of the basis-dependent and basis-independent measures of coherence for a single qubit. 
For the basis-dependent coherence, the set of incoherent states lie on the $ z $-axis, and the coherence is defined as
the minimal distance to this line.  For the basis-independent coherence, the reference state is the maximally mixed state
$ \rho_I = I/2 $.  }
\label{fig:bloch}
\end{figure}

The above relationship can be neatly illustrated for a single qubit using the Bloch sphere representation (see Fig. \ref{fig:bloch}).  
In the basis-dependent definition of coherence, the coherence is measured from the quantum state to the nearest point on a line 
passing through the center of the Bloch sphere.  In the case of coherence measured in the $ z $-basis, this correspond to the distance 
from the state to the $ z $-axis. Meanwhile, in the basis-independent definition, the coherence is defined to be the 
distance to the center of the Bloch sphere. In the case of a multipartitite system,  the basis-dependent approach 
measures the distance in a $d^2 - 1$ dimensional space to the closest incoherent 
state in the hyperplane of dimension $d-1$. In the basis-independent approach the distance is measured from the original state 
to a unique maximally mixed state $ \rho_I $ within the $d^2 - 1$ dimensional space.

\section{Distribution of coherence in a multipartite system}
\label{sec:dist}

\subsection{Basis-independent intrinsic and local coherence}

The quantum coherence as defined in (\ref{coherencedefinition}) and (\ref{coherencemeasure})  
measures the total coherence in the system.  In Ref. \cite{radhakrishnan2016distribution}, we showed that the coherence can be distributed in various ways in a multipartite system.  The quantum coherence may exist  due to the collective participation of several subsystems, or can be attributed to coherence located within the subsystems.  
The former type of coherence is called the {\it intrinsic coherence} and defined as 
\begin{equation}
C_I (\rho)  \equiv \underset{\sigma_S \in S}{\hbox{min}}  
{\cal D} (\rho,\sigma_S),
\label{intrinsicdef}
\end{equation}
where the minimization is performed over all separable states
\begin{align}
\sigma_S = \sum_k p_k \sigma_1^{(k)} \otimes \sigma_2^{(k)} \otimes \dots \otimes  \sigma_N^{(k)}
\label{separablestate}
\end{align}
where $ \sigma_i $ is a density matrix on the $ i $th subsystem, and $ N $ is the total number of subsystems. This type of coherence measures the quantum coherence which arises due to the non-local quantum correlations, and is formally equivalent to the entanglement in the system. The terminology of ``intrinsic'' coherence arises because it is a coherence which cannot be removed by a local basis transformation. This can be viewed as the contribution of the entanglement to the coherence.  The remaining part of the coherence is called the {\it local coherence} and is defined as
\begin{equation}
C_L^{(b)} (\rho) \equiv {\cal D} (\sigma_S^{\min} , \rho_d^{(b)} ),
\label{localcoherence}
\end{equation}
where $ \sigma_S^{\min} $ is the minimal value from the optimization of (\ref{intrinsicdef}), and $ \rho_d^{(b)} $ is the same matrix as $ \rho $ with all off-diagonal elements set to zero:
\begin{align}
\langle b_n | \rho_d^{(b)} | b_m \rangle  = \delta_{nm} \langle b_n | \rho | b_m \rangle ,
\label{rhoddef}
\end{align}
and the states $ | b_n \rangle $ are a set of orthonormal states.  Since the state $ \rho_d^{(b)} $ depends upon the chosen basis specified by the 
states $ | b_n \rangle $, the local coherence is a basis dependent quantity.  The intuition of calling (\ref{localcoherence}) the ``local'' coherence is that it is the 
coherence contained with the separable state (\ref{separablestate}). The coherence arises due to the local contributions within 
the separable state.  Another way to view this is that the entanglement gives a non-local contribution to the coherence, 
while the remaining is the local coherence. 

In a basis-independent definition of coherence, instead of using $ \rho_d^{(b)} $ as the reference state, the completely mixed state $ \rho_I $ is used.  The intrinsic coherence $ C_I  $ is unchanged since it does not involve $ \rho_d^{(b)} $, but the basis-independent local coherence is written
\begin{align}
C_L (\rho) & \equiv {\cal D} (\sigma_S^{\min}, \rho_I ) \nonumber \\
& =  \sqrt{S \left( \frac{\sigma_S^{\min} +I/d}{2}\right)  
               - \frac{S(\sigma_S^{\min}) + \log_{2} d }{2} }.
\label{localcoherencebi}
\end{align}

The intrinsic coherence is invariant under local unitary transformations
\begin{align}
C_I ( U_{\text{loc}}^\dagger \rho U_{\text{loc}}) = C_I (\rho)
\end{align}
where 
\begin{align}
U_{\text{loc}} = U_1 \otimes U_2 \otimes \dots \otimes U_N 
\end{align}
and  $ U_i $ is a unitary transformation on the $ i$th subsystem. This can be shown to be true in the usual way that 
entanglement is invariant under local transformation.  The local coherence is similarly invariant under local transformations
\begin{align}
C_L  ( U_{\text{loc}}^\dagger \rho U_{\text{loc}}) = C_L (\rho) 
\end{align}
since it is implicitly dependent upon the minimized state $ \sigma_S^{\min} $.  These relations are more restrictive than the total coherence 
because it involves the intermediate state $  \sigma_S^{\min} $.  This is because a general unitary transformation that is considered in (\ref{generalcoherence}) does not preserve the structure of the subsystems, as we are interested in doing by decomposing the coherence into its two parts. 

We may relate the basis-independent intrinsic and local coherence to the total coherence (\ref{coherencemeasure}) in a similar way to that argued 
in Ref. \cite{radhakrishnan2016distribution} for the basis-dependent case.  Due to the metric property of QJSD and the triangle inequality, we have the relation
\begin{align}
C (\rho) \le C_I(\rho)+ C_L(\rho)  .
\label{originaldecomp}
\end{align}
This relation will allow us to characterize the contributions to coherence in a multipartite system.

\subsection{Alternative decompositions of coherence}

The above decomposition is not the only possibility for identifying the contributions to coherence 
in a multipartite system.  The definition (\ref{intrinsicdef}) emphasizes the contribution of 
the entanglement to the coherence.  However this is merely one choice,
and other definitions which show other types of correlations can be equally used.  By changing the 
intermediate state from $ \sigma_S^{\min} $ to another state, one can create variants of the same basic idea.  
An alternative is to use the product state 
\begin{align}
\pi =  \sigma_{1} \otimes \cdots \otimes \sigma_{N} .
\label{productstate}
\end{align}
Using this as the intermediate state, we define the {\it collective coherence} to be
\begin{align}
C_{c}(\rho) & =  \underset{\pi \in \mathcal{I}_P}{\hbox{min}} 
{\cal D}  (\rho,\pi)  
\label{globalcoherencedef}        
\end{align}
where $  \mathcal{I}_P $ is the set of all product states of the form (\ref{productstate}). For the QSJD, we show in the Appendix that the closest product state is 
\begin{align}
\pi^{\text{min}} = \pi_{\rho} \equiv \rho_1 \otimes \cdots \otimes \rho_{N} 
\label{piminimum}
\end{align}
where $ \rho_i $ is the reduced density matrix for the $ i$th subsystem.  Hence the collective coherence is in our case
\begin{align}
C_{c}(\rho)  = \sqrt{S \left( \frac{ \rho + \pi_{\rho} }{2}\right)  
               -\frac{S(\rho)+ S(\pi_{\rho}  )}{2}}. 
\label{globalcoherence}               
\end{align}
Since the minimization can be evaluated explicitly, this makes (\ref{globalcoherence}) a much more efficient decomposition from a practical point of view since finding the closest separable state (\ref{intrinsicdef}) is typically a highly numerically 
intensive procedure.

The state (\ref{productstate}) differs from (\ref{separablestate}) in that all correlations between the 
subsystems are completely broken.  As discussed in Ref. \cite{modi2010unified}, the distance from a quantum state to the product state 
is equal to the total mutual information between the systems, including both quantum and classical correlations between 
the two subsystems.   We thus define the {\it localized coherence} of the system as the QJSD distance from the product state $\pi(\rho)$ to the maximally mixed state $\rho_{I}$
\begin{align}
C_{l}(\rho) & = {\cal D}  (\pi^{\text{min}},\rho_I) \nonumber \\
& =  \sqrt{S \left( \frac{\pi_{\rho} +I/d}{2}\right)  
               - \frac{S(\pi_{\rho}) +\log_{2} d}{2} }.
\label{localizedcoherence}               
\end{align}
By measuring the distance between the product and the maximally incoherent state, this corresponds to finding only the part of coherence which is localized within the qubits. 

The collective and localized coherence obey the same invariance properties as the coherences of the previous section under local unitary transformations.  As before
\begin{align}
C_c ( U_{\text{loc}}^\dagger \rho U_{\text{loc}}) & = C_c (\rho) \nonumber \\
C_l ( U_{\text{loc}}^\dagger \rho U_{\text{loc}}) & = C_l (\rho)
\label{localunitary}
\end{align}
which follows from the fact that the form of the product states (\ref{productstate}) are preserved under such local transformations.  Similarly to (\ref{originaldecomp})  the these coherences can be decomposed as
\begin{equation}
C(\rho)  \leq C_{c}(\rho)  + C_{l} (\rho) .
\label{triangleinequality}
\end{equation}
Furthermore, we can deduce that 
\begin{align}
C_I(\rho)  \le C_c(\rho) 
\label{cicc}
\end{align}
from the fact that the product state (\ref{productstate}) is a special instance of the more general class of separable states (\ref{separablestate}).  We can understand (\ref{cicc}) from the fact that the collective coherence $ C_c $ contains contributions from all correlations (quantum and classical), where as $ C_I $ only contains contributions from entanglement, a more specialized type of correlation.

\begin{figure}[t]
\includegraphics[width=\columnwidth]{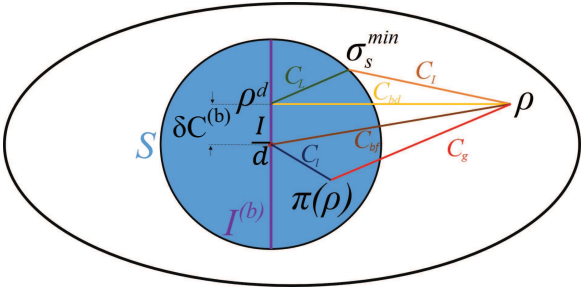}
\caption{Definition of the various coherences as the distance between the states. $ S $ is the 
set of all separable states and $I^{(b)}$ is the set of all incoherent states and $I/d$ is the maximally
mixed state. $\rho_{d}$ and $\sigma^{min}_{s}$ are the closest incoherent state and the closest 
separable state found using the coherence estimation method in Ref. \cite{radhakrishnan2016distribution}
and $\pi(\rho)$ is the product state of the density matrix $\rho$. }
\label{fig2}
\end{figure}

It may appear strange that ``classical correlations'' should contribute to the collective coherence, or to any kind of coherence at all.  The issue here is mainly due to nomenclature of what is deemed to be classical or quantum.  In the context of {\it correlations}, a state such as 
\begin{align}
\frac{1}{2} \left( |00 \rangle \langle 00 | + |11\rangle \langle 11|  \right)
\label{example0011}
\end{align}
is considered ``classical'' since it can be written perfectly using a classical probability distribution.  In the context of {\it coherence}, there are bases where such a state contains coherence, for instance in the $ |\pm \rangle $ basis, the same state is written
\begin{align}
\frac{1}{4} \left(
\begin{array}{cccc}
1 & 0 & 0 & 1 \\
0 & 1 & 1 & 0 \\
0 & 1 & 1 & 0 \\
1 & 0 & 0 & 1
\end{array}
\right) .
\label{pmcorrelated}
\end{align}
This state contains off-diagonal components and therefore possesses coherence. For the basis-dependent approach, a state such as (\ref{example0011}) may or may not be considered to possess coherence depending on the chosen basis.  On the other hand, in the basis-independent approach, the state
 (\ref{example0011})  always has coherence because it is distinct to the maximally mixed state $ \rho_I $.

We have introduced several ways of defining the coherence using a basis-dependent and basis-independent reference state, and also two options for decomposing the coherence.  To understand the connection between these quantities, we can use a geometrical view to contrast the relationship between them.  In Fig. \ref{fig2} we represent the Hilbert space of an arbitrary quantum system showing the set of all separable states $ S $, the set of incoherent states  $  \mathcal{I}^{(b)} $, and the maximally mixed state $ \rho_I = I/d$. The maximally mixed state lies on the line of incoherent states as it is always contained in  $  \mathcal{I}^{(b)} $, regardless of the basis chosen.  Different basis $ b $ will correspond to different hyperplanes (represented by a line in Fig. \ref{fig2}) throughout the Hilbert space.   The lines connecting the density matrices depict the coherences in the system.  We see that the two decompositions (\ref{originaldecomp}) and (\ref{triangleinequality}) can be visualized as two triangles sharing one corner.  The total coherence in the case of the basis-dependent approach is the closest state in the set $  \mathcal{I}^{(b)} $.  Meanwhile the total coherence in the basis-independent case is the line to the point $ \rho_I = I/d$.  Each of these two can be decomposed into two contributions according to the triangle inequality depending upon whether the separable state $ \sigma_S $ is used or the product state $ \pi $. The relation (\ref{cicc}) follows from the fact that a product state is particular form of a separable state, and will generally have a larger value.

We note that other possibilities of intermediate states also can be chosen.  For instance, to restrict the correlations to purely quantum correlations one may also define coherences based on a quantum discord, where the intermediate state is the closest state to the form
\begin{align}
\chi = \sum_{k_1, k_2, \dots, k_N} p_{k_1 k_2 \dots k_N} |k_1 k_2 \dots k_N \rangle \langle k_1 k_2 \dots k_N | ,
\label{classicallycorrelated}
\end{align}
where the states $ | k_i \rangle $ are orthogonal states (in an arbitrary local basis).  While we do not explicitly define coherences based on this intermediate state, similar properties to what we have shown above follow, and give different contributions to the coherences.

\subsection{Coherence due to basis choice}

From Fig. \ref{fig2} we can see that the
basis-dependent coherence and the basis-independent coherence can be connected through a new line 
joining the points $\rho_{d}$ and the maximally mixed state $I_{d}/d$.  This quantity 
can be interpreted as the additional contribution to the coherence in the basis-independent approach as 
compared to the basis-dependent coherence:
\begin{align}
\delta C^{(b)} (\rho)  & = {\cal D}  (\rho_{d},\rho_{I})  \nonumber  \\
& = \sqrt{S \left( \frac{\rho_{d} +I/d}{2}\right) - 
\frac{ S(\rho_{d})+ \log_{2} d}{2}       }.
\label{coherencelost}                
\end{align}
Here $\rho_{d}$ is the closest incoherent state which is obtained during the optimization of 
(\ref{coherencedefinition}).  Here we find that the three forms of coherences given through equations 
(\ref{coherencedefinition}),
(\ref{coherencemeasure}) and (\ref{coherencelost}) are geometrically interrelated and 
obey the triangle inequality  
\begin{equation}
C (\rho)  \leq C^{(b)} (\rho)  + \delta C^{(b)} (\rho).
\label{triangleinequality2}
\end{equation} 
Using the basis-dependent version of the inequality (\ref{triangleinequality}), we can also relate the 
basis-independent coherence to the basis-dependent quantities:
\begin{eqnarray}
C (\rho)  & \leq & C_{I}(\rho)  + C_{L}^{(b)} (\rho)  + \delta C^{(b)}  (\rho).
\label{triangleinequality3}
\end{eqnarray}
From Fig. \ref{fig2} we can see that (\ref{triangleinequality3}) describes 
a quadrilateral inequality defined by the points $\rho$, $\sigma_{s}^{\min}$, $\rho^{d}$ and $I_{d}$. 
Under the conditions that $\sigma_{s}^{\min} = \rho^{d} = I_{d}$, the area of the quadrilateral
is zero and hence $C(\rho) = C_{I}(\rho)$.

\section{Examples}
\label{qcmpml}

\subsection{Two site transverse Ising model}

We now show some examples of the basis-independent coherence and its decomposition in multipartite systems.   
One of the fundamental properties of coherence that was discussed in Ref. \cite{radhakrishnan2016distribution} was the trade-off in coherence between the local and non-local (i.e. intrinsic) coherences.  Our first example examines the two site transverse Ising model using the collective and localized coherence as defined in (\ref{globalcoherence}) and (\ref{localizedcoherence}). The Hamiltonian is
\begin{align}
H = -2 \sin \xi \sigma_1^z \sigma_2^z - \cos \xi  ( \sigma^x_1 + \sigma^x_2 )  ,
\end{align}
which can be diagonalized to give the  ground state 
\begin{align}
|E_0 \rangle =& \frac{1}{\sqrt{\cal N}} \Big[  ( 1 - \sin \xi) (  | 0 1 \rangle + | 10 \rangle) \nonumber \\
& + \cos \xi ( | 0 0 \rangle + | 11 \rangle ) \Big]
\label{groundising}
\end{align}
where $ {\cal N} $ is a suitable normalization factor.  The ground state has the limiting behavior
\begin{align}
|E_0 \rangle = \left\{
\begin{array}{cc}
|++ \rangle & \hspace{1cm} \xi \rightarrow 0 \\
 \frac{1}{\sqrt{2}} (| 0 0 \rangle + | 11 \rangle ) & \hspace{1cm} \xi \rightarrow \pi/2 
\end{array}
\right.
\end{align}
where $ |+ \rangle = (|0 \rangle + | 1 \rangle )/\sqrt{2} $.  

\begin{figure*}[t]
\includegraphics[width=\textwidth]{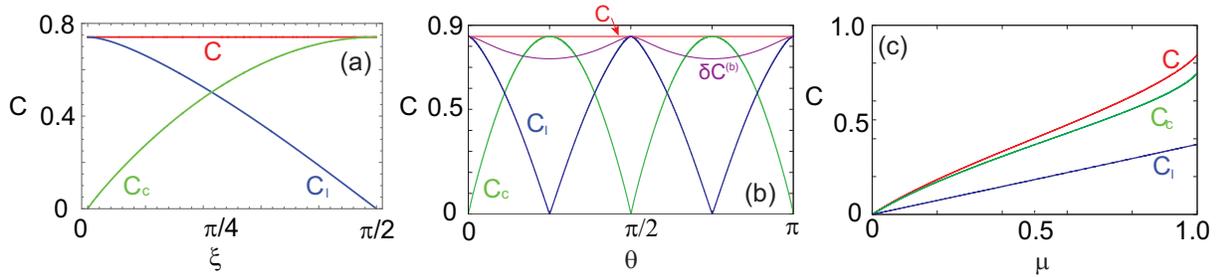}
\caption{Quantum coherence and distribution for (a) the ground state of the transverse Ising model and 
(b),(c) the generalized GHZ state.  
For the GHZ case, the coherences for the case of (b) pure states $ \mu = 1 $, 
(c) mixed states with $ \theta = \pi/3 $.   } 
\label{fig3}
\end{figure*}

Figure \ref{fig3}(a) shows the collective and localized coherence calculated using (\ref{globalcoherence}) and (\ref{localizedcoherence}), with the product state taken as being the tensor product of the reduced density matrices for the state (\ref{groundising}).  We see a similar trade-off of the coherence as was observed in Ref. \cite{radhakrishnan2016distribution}.  This can be easily understood since the only contribution of the coherence in a Bell state ($ \xi \rightarrow \pi/2  $) is due to the collective nature of the entanglement.  Meanwhile, in the other limit the completely separable $|++\rangle$ state has all the coherence localized in the two qubits. 

One difference to the previously obtained result using the basis-dependent coherence  (Fig. 2(a) of Ref. \cite{radhakrishnan2016distribution}) is that the total coherence was different in the two limits, whereas in Fig. \ref{fig3}(a) the total coherence is a constant for all $ \xi $.  The reason for this is that as discussed in (\ref{triangleinequality2}) there is an additional contribution $ \delta C^{(b)} $ to the basis-independent coherence due to the different definition of the incoherent state.  Furthermore, the fact that the total coherence is a constant is guaranteed for the state (\ref{groundising}) since it is a pure state for any value of $ \xi $. Our numerically calculated value agrees with the formula (\ref{totalcoherenceformula}) for $ d = 4 $.  

We find that the basis-independent procedure tends to be a more robust procedure for performing the coherence decomposition.  This occurs in situations where there is no unique solution to the minimization procedure in (\ref{intrinsicdef}).  For instance, for the Bell states, the closest separable state is not unique due to the property that one can write
\begin{align}
 \frac{1}{\sqrt{2}}\left(| 0  0 \rangle + | 1  1 \rangle \right)=  \frac{1}{\sqrt{2}}\left( | \theta  \theta \rangle + | \bar{\theta} \bar{\theta}  \rangle  \right)
\end{align}
where $ | \theta \rangle = \cos \theta | 0 \rangle + \sin | 1 \rangle $ and $ | \bar{\theta} \rangle = \sin \theta | 0 \rangle - \cos | 1 \rangle $.  The closest separable state in this case is 
\begin{align}
\frac{1}{2}  \left( | \theta  \theta \rangle \langle \theta \theta | + | \bar{\theta}  \bar{\theta} \rangle \langle \bar{\theta}  \bar{\theta} | \right)
\end{align}
which can take any value of $ \theta $.  Thus depending upon the particular $ \sigma_S^{\text{min}} $ that is found, one may obtain different values of the local coherence in the basis-dependent procedure, in a similar way to (\ref{pmcorrelated}).  However, in the basis-independent approach, thanks to the properties (\ref{localunitary}) if the solutions are related by a local unitary transformation then the obtained coherence is identical.  This is an advantage of the basis-independent approach from a practical point of view of application of the theory.

\subsection{Tripartite systems}

We now turn to investigating the basis-independent quantum coherence for tripartite 
systems. Based on how they are entangled, a three qubit system can be classified into 
two different classes: the GHZ and W class of 
states.  These two different classes of states cannot be connected to each other via local operations and classical communications (LOCC)
\cite{dur2000three}. To have a complete understanding of the distribution of quantum coherence we will study 
these two classes of states.

The generalized GHZ state is defined as 
\begin{align}
|\text{GHZ} \rangle = \cos \theta |000 \rangle + \sin \theta |111 \rangle; \; \theta \in (0,2\pi],
\end{align}
from which we evaluate the total, collective, and local coherence. Our results are shown in Fig. \ref{fig3}(b).  We again find that the total coherence is invariant of $ \theta $ for the same reasons as discussed in the context of Eq. (\ref{totalcoherenceformula}).  
The local and global coherence exhibit an oscillatory trade-off behavior, as observed for the two qubit example.  The collective coherence
is a maximum where the localized coherence is a minimum, and vice versa. This again has the simple interpretation that the coherence 
contributions change depending on the parameter $ \theta $, which controls the degree of correlations between the qubits. 
We also verify that the sum of the collective and localized coherence is always greater than total coherence, obeying the triangle inequality (\ref{triangleinequality}) as expected. As with the two qubit example in Fig. \ref{fig3}(a), if one were to use the basis-dependent measure, the total coherence would not be constant with $ \theta $. The extra contribution to the coherence  $\delta C^{(b)} $  due to the basis-independent measure is shown also in Fig. \ref{fig3}(b).  We verify the triangle inequality (\ref{triangleinequality2}) for the generalized GHZ states. 

To consider mixed states, we consider states of the Werner type (\ref{wernertype}) 
where the state $ |\psi \rangle = |\text{GHZ} \rangle $ with $\theta=\pi/3$.  
The effect of mixing 
is illustrated in Fig. \ref{fig3}(c).  We observe that the total, collective, and localized coherences are all monotonically dependent on the
mixing parameter, as the state approaches the maximally incoherent state which by definition has zero coherence.  
We verify that the triangle inequalities (\ref{triangleinequality}) and (\ref{triangleinequality2}) are obeyed for all values of the mixing parameter.  All these show the expected behavior for a basis-independent coherence measure and its decompositions.

\begin{figure*}[t]
\includegraphics[width=\textwidth]{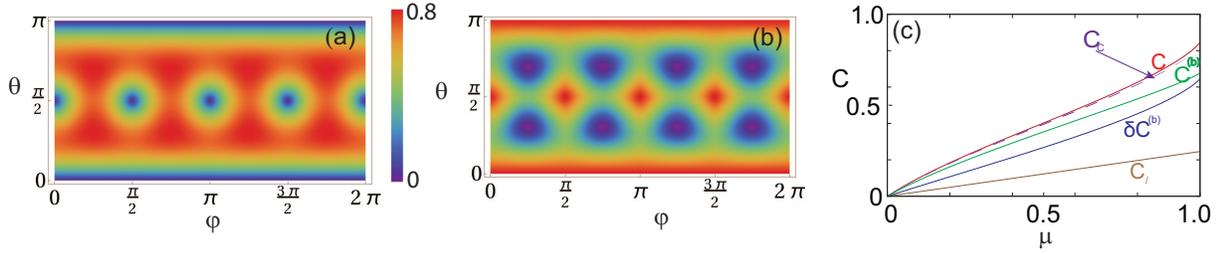}
\caption{Quantum coherence and distribution for the W state (\ref{wstatedef}). The (a) collective and (b) localized coherence is shown for pure states $ \mu = 1 $. (c) The dependence of various coherences with the mixing parameter $ \mu $ for  $ \sin \theta \sin \phi  =  \sin \theta \cos \phi = \cos \theta = 1/\sqrt{3} $.  } 
\label{fig4}
\end{figure*}

Similar results are obtained for the W class of states.  We define the generalized W state as
\begin{align}
|\text{W} \rangle = \sin \theta \sin \phi |001 \rangle
+ \sin \theta \cos \phi |010 \rangle + \cos \theta |100 \rangle ,
\label{wstatedef}
\end{align}
where $\theta \in (0,\pi]$ and $\phi \in (0,2 \pi]$.
The total coherence is found to give a constant $ C \approx 0.847 $ for all parameters and is consistent with (\ref{totalcoherenceformula}).  We show the collective and localized coherence for the pure state case in Fig.  \ref{fig4}(a)(b).  We again see a trade-off behavior where the localized coherence is a maximum when the collective coherence is a minimum and vice versa.  For mixed states, we use again a Werner form (\ref{wernertype}) where $ | \psi \rangle = |\text{W} \rangle  $. Similar to the GHZ state, we observe that all coherences depend monotonically with the mixing parameter as expected. Triangle inequalities for the collective and localized coherences, as well as the difference in the total coherence due to basis-independent and basis-dependent measures were verified and shown to be true for all parameters.

\section{Summary and conclusions}
\label{sumconc}

In this paper, we analyzed quantum coherence in a basis-independent manner, by using the maximally mixed state as
the reference incoherent state.  This is in contrast to the earlier studies where coherence
was studied as a property with reference to a particular chosen basis $ b $.  This is an attractive formulation for 
quantifying coherence as it gives an objective property to a quantum state, and gives desirable characteristics such 
as invariance under basis transformations.  This approach to quantifying coherence is closely related to the purity of the system, whereas the original basis-dependent approach is quantifying the superposition with respect 
to a particular basis.  Apart from conceptual advantages of giving an objective quantification of a state independent of the observer's choice, it also gives simpler and more robust procedure that is free of optimization.  

The approach naturally can be applied to coherence decomposition techniques, where the 
distribution of the coherence in a multipartite system can be found.  We introduced an 
alternative decomposition where the coherence is split into the collective and localized 
contributions.  As this method only requires calculation of the product state of the system, 
it is a simple yet effective way of performing the coherence decomposition. The decomposed coherences
obey invariance properties under local unitary transformations.  
We verified that there is a coherence trade-off between these quantities in a variety of systems.  
These trade-off relations are 
captured thanks to the triangle inequality which is a very fundamental property of the 
square root of the Jensen-Shannon divergence. For other coherence measures, we expect that similar
relations hold, although the inequality relations derived in this paper should only be strictly true 
for the measure that we have chosen here.  

The use of local and collective strategies has been carried out in quantum metrology 
\cite{higgins2007entanglement,sahota2015quantum,knott2016local,ilo2014theory} to gain 
interferometric advantages. Further it has been shown that in quantum 
computation protocols, collective coherence provides the necessary resource for an exponential 
computational speed-up compared to classical algorithms \cite{shahandeh2017quantum,shahandeh2017quantum2,byrnes2015macroscopic,pyrkov2014quantum}.  Although these factors show the resource theoretic value of quantum coherence, a simple and effective basis-independent computation of the local and collective coherences is highly valuable to have a estimate of the available resource. This may be one of the practical applications of 
our formalism.

\section*{Acknowledgements} 
TB is supported by the Shanghai Research Challenge Fund; 
New York University Global Seed Grants for Collaborative Research; 
National Natural Science Foundation of China (61571301); 
the Thousand Talents Program for Distinguished Young Scholars (D1210036A); 
and the NSFC Research Fund for International Young Scientists (11650110425); 
NYU-ECNU Institute of Physics at NYU Shanghai; 
the Science and Technology Commission of Shanghai Municipality (17ZR1443600); 
and the China Science and Technology Exchange Center (NGA-16-001).

\appendix

\section{Proof of the closest product state for the quantum Jensen-Shannon divergence}

We use a similar proof to that given in Ref. \cite{modi2010unified} to show that the closest product state for an 
arbitrary density matrix is the tensor product of the reduced density matrices (\ref{piminimum}).  First assume
that the state 
\begin{align}
\alpha = \alpha_1 \otimes \dots \otimes \alpha_N
\end{align}
is the closest product state to $ \rho $.  This assumption implies that 
\begin{equation}
{\cal D} ( \rho, \pi_\rho ) - {\cal D} ( \rho, \alpha)  \ge 0 ,
\end{equation}
which for the QJSD can be written
\begin{align}
\sqrt{{\cal J} ( \rho, \pi_\rho )} - \sqrt{{\cal J} ( \rho, \alpha)} \ge 0 
\label{alphaassump}
\end{align}
From the triangle inequality of the QJSD we have
\begin{align}
\sqrt{{\cal J} ( \rho, \alpha )} \ge \sqrt{{\cal J} ( \rho, \pi_\rho )} + \sqrt{{\cal J} ( \alpha , \pi_\rho )} .
\end{align}
Rearranging this inequality and combining this with (\ref{alphaassump}) gives
\begin{align}
- \sqrt{{\cal J} ( \alpha , \pi_\rho )} \ge  \sqrt{{\cal J} ( \rho, \pi_\rho )}  - \sqrt{{\cal J} ( \rho, \alpha )} \ge 0  .
\end{align}
The quantity $ - \sqrt{{\cal J} ( \alpha , \pi_\rho )} $ is a negative quantity with equality only when $ \alpha =  \pi_\rho $.  Therefore, for all states $ \rho $ we find 
\begin{align}
 \underset{\pi \in \mathcal{I}_P}{\hbox{min}} \sqrt{{\cal J}  (\rho,\pi)  } = \sqrt{{\cal J}  (\rho ,\pi_\rho)  }  .
\end{align}
Thus we prove that for the Jensen-Shannon divergence, the closest product state is the 
one composed of the single qubit reduced density matrices.


\end{document}